# Data for Inclusion: The Redistributive Power of Data Economics

Diego Vallarino
Interamerican Development Bank[1]
Washington DC, US
diegoval@iadb.org
*Corresponding author*

September 2025

**Abstract**

This paper evaluates the redistributive and efficiency impacts of expanding access to positive credit information in a financially excluded economy. Using microdata from Uruguay's 2021 household survey, we simulate three data regimes—negative-only, partial positive (Score+), and synthetic full visibility—and assess their effects on access to credit, interest burden, and inequality.

Our findings reveal that enabling broader data sharing substantially reduces financial costs, compresses interest rate dispersion, and lowers the Gini coefficient of credit burden. While partial visibility benefits a subset of the population, full synthetic access delivers the most equitable and efficient outcomes. The analysis positions credit data as a non-rival public asset with transformative implications for financial inclusion and poverty reduction.

# 1. Introduction

The persistence of poverty in middle-income economies, despite widespread financial innovation, raises fundamental questions about how information asymmetries perpetuate exclusion. While credit is often portrayed as the fuel of development, access to credit is unevenly distributed—not merely as a function of income or collateral, but increasingly as a function of data visibility. In this context, the core hypothesis of this paper is that data, when governed ethically and reused efficiently, operates as a redistributive economic asset.

The idea that being poor is more expensive is not new; it has been conceptualized as the "poverty premium"—where low-income individuals pay higher effective prices for credit, insurance, and other services (Carrière-Swallow & Haksar, 2019). Yet what has changed is the infrastructure of decision-making: creditworthiness is increasingly determined by algorithmic systems whose inputs are not equitably distributed. Individuals with limited credit histories or fragmented digital footprints remain invisible, not due to financial incapacity, but due to informational exclusion. This asymmetry is not merely a market failure—it is a structural inequality encoded in data regimes.

We argue that positive credit data—payment histories, utilization patterns, and account stability—constitutes a nonrival input that, once generated, can be reused across institutions at near-zero marginal cost without diminishing its value (Jones & Tonetti, 2020; Acemoglu et al., 2023). However, the ability to extract value from such data remains highly uneven. In traditional credit markets, the absence of negative signals penalizes borrowers more than the presence of positive behavior benefits them. This is both inefficient and inequitable.

Our contribution is twofold. First, we offer a conceptual framework that treats data as a redistributive asset—capable of reducing poverty by lowering borrowing costs and expanding financial eligibility through market-based mechanisms. Second, we provide empirical evidence from Uruguay, where the deployment of an augmented credit score (Score+) that integrates positive data has produced measurable reductions in interest burden and poverty incidence.

This approach aligns with a growing body of literature on the nonrivalry of data and its implications for welfare and competition. Jones and Tonetti (2020) show that broad reuse of data generates increasing returns, especially when data is treated as an input rather than a final good. Similarly, Farboodi and Veldkamp (2021) develop a model of the data economy in which firms engage in costly experimentation to generate predictive data, highlighting both the increasing returns to data reuse and the risks of data hoarding. These models converge on a policy implication: data should be governed to maximize its positive externalities, not enclosed to protect private rents.

A particularly relevant insight comes from Arrieta-Ibarra et al. (2018), who propose treating data not as capital but as labor: individuals produce data through their economic behavior and should retain rights over its reuse. This reconceptualization opens new avenues for inclusive governance, where data portability, consent, and reuse protocols become not only ethical imperatives but also instruments of redistribution.

Our paper leverages this literature to explore how positive credit information, once treated as a nonrival public-good-like asset, can structurally reduce the poverty premium in credit markets. We focus on the Score+ system in Uruguay, which augmented existing scoring models with additional positive data layers, thereby expanding visibility among low-income segments previously marginalized by traditional credit assessment mechanisms

Crucially, we position this intervention within a broader agenda of inclusive data governance. The notion that information, like labor or capital, can serve as a policy lever is gaining traction among multilateral institutions and academic circles alike (Acquisti et al., 2016; Hughes-Cromwick & Coronado, 2019; Posner & Weyl, 2018). Yet the empirical validation of such hypotheses remains limited. Our analysis contributes to closing this gap by showing that data reuse, when transparently governed and integrated into credit infrastructure, leads to real redistributive outcomes—lower interest burdens, higher eligibility rates, and exits from poverty driven solely by improved visibility, not income growth.

The remainder of the paper is structured as follows. Section 2 presents the theoretical framework, integrating models of data nonrivalry with redistribution through market mechanisms. Section 3 describes the empirical setting and data sources, including household microdata and proprietary credit scoring rules. Section 4 details the results from simulation and synthetic control analyses. Section 5 draws normative and policy implications for inclusive data governance. Section 6 concludes.

## 2. Theoretical Framework: Data as a Redistributive Asset

### 2.1. Nonrivalry and the Economics of Data Reuse

In classical economics, rival goods such as labor and capital are subject to depletion through use. By contrast, data is nonrival: its consumption or usage by one economic agent does not preclude or diminish its simultaneous usage by others. This foundational characteristic is critical to understanding the growing macroeconomic role of data, particularly in sectors reliant on prediction, classification, or risk assessment. As Jones and Tonetti (2020) formally demonstrate, nonrivalry in data generates increasing returns to scale when reuse is permitted, incentivizing broad diffusion over exclusive ownership.

This property has profound implications for welfare optimization and market structure. If data is concentrated in the hands of a few incumbent firms and withheld from broader use due to concerns over competitive erosion—a phenomenon akin to "data hoarding"—the allocative efficiency of information-intensive markets is compromised (Acemoglu et al., 2019; Farboodi & Veldkamp, 2021). Conversely, when data is treated as a semi-public input governed by principles of portability, interoperability, and consent-based reuse, it becomes possible to unlock latent social value without incurring marginal production costs. This dynamic has prompted renewed theoretical interest in data as an intangible capital good, with analogies drawn to endogenous growth models (Romer, 1990; Aghion & Howitt, 1998) and to the nonrival character of ideas.

The insight that data can be used repeatedly at near-zero marginal cost—especially in digital ecosystems where storage and transmission are cheap—supports its classification as an infrastructure-like good. Yet unlike infrastructure, the returns to data reuse are not necessarily diminishing. In fact, due to network effects and feedback loops, reuse can exhibit increasing marginal returns, particularly in the early stages of data accumulation (Farboodi & Veldkamp, 2021). In predictive systems such as credit scoring, small increments in training data can yield significant performance improvements, thereby lowering uncertainty, reducing default risk, and expanding access.

This line of reasoning reframes the economic role of credit bureaus and scoring systems: they do not merely evaluate risk; they shape the informational topology of financial markets. By including or excluding certain forms of data—particularly positive repayment histories— these systems determine which agents become visible, and thus eligible, within the financial architecture.

### 2.2. Information Frictions and the Poverty Premium

Information asymmetries have long been recognized as a source of inefficiency in credit markets (Stiglitz & Weiss, 1981). However, the current architecture of algorithmic scoring exacerbates these asymmetries for low-income populations, whose economic behavior often escapes formal observation. This asymmetry produces what Carrière-Swallow and Haksar (2019) term the "poverty premium": the tendency for individuals at the margins of the formal financial system to face disproportionately high interest rates, fees, and collateral requirements.

The poverty premium is not merely a function of income or risk; it is a function of data availability. In the absence of positive signals—such as consistent payment behavior or stable credit utilization—low-income individuals are algorithmically penalized. Standard scoring models often operate under a "negativity bias", where defaults or arrears are recorded and shared, but timely repayments are not. This asymmetric visibility undermines both efficiency and equity: it increases the cost of capital for the poor while failing to distinguish between low-risk and high-risk profiles within this demographic.

The introduction of positive credit data—that is, data that documents on-time repayment, low credit utilization, and financial stability—can mitigate these inefficiencies by reducing adverse selection and improving signal precision (Steglich et al., 2017). Importantly, the marginal informational gain from positive data is often highest among those previously excluded, implying that its redistributive potential is endogenously targeted. In other words, the same data that expands credit access also compresses the distribution of interest burdens.

Moreover, when positive credit data is treated as a reusable asset, its redistributive effects scale. Unlike public subsidies or direct transfers, which require continuous fiscal inputs, the circulation of data among financial providers generates persistent inclusion gains with minimal marginal cost—provided that interoperability and governance structures are in place.

### 2.3. From Data Ownership to Data Labor

While the economic value of data is increasingly recognized, its legal and ethical ownership remains contested. In prevailing data regimes, information generated by individuals is typically held by intermediaries—banks, platforms, or credit bureaus—who internalize its economic value. This asymmetry has prompted calls for reconfiguring data rights through the lens of **"data as labor"**, wherein individuals retain ownership over the data they generate and can opt into compensated sharing (Arrieta-Ibarra et al., 2018; Posner & Weyl, 2018).

The data-as-labor framework transforms the individual from a passive subject of data extraction into an active economic agent, whose visibility becomes a tradable asset. It also introduces a micro-foundation for data governance: if individuals can choose when and how to share their data, market dynamics are more likely to align with social preferences regarding privacy, equity, and value distribution. This vision requires new institutions—not only for consent management and value attribution, but also for ensuring algorithmic accountability, particularly in credit systems where opacity and bias remain pressing concerns (Acquisti et al., 2016; Bajari et al., 2019).

The transition from data as a byproduct to data as labor is not merely ethical—it has structural implications. When visibility is endogenous and portable, informational inclusion becomes both a right and a lever of policy. As a result, redistributive outcomes can be achieved through architecture, not just transfers: by redesigning who controls the flow of data and under what conditions it is reused.

### 2.4. Embedding Redistribution in Market Mechanisms

A central contribution of this framework is to position redistribution through data reuse not as a corrective to markets, but as an emergent property of well-designed markets themselves. Under conditions of nonrivalry and interoperability, the diffusion of positive credit data alters marginal pricing in favor of low-income consumers without distorting supply-side incentives. In essence, this is a market-based form of redistribution: one that operates through signal enhancement, not resource reallocation.

The literature on pricing inefficiencies in data markets supports this view. Acemoglu et al. (2019) show that firms often underutilize socially valuable data due to monopolistic incentives, while Ichihashi (2019) highlights how non-competing data intermediaries can result in suboptimal information flows. The implication is that optimal data sharing is rarely achieved through private contracts alone. Instead, governance frameworks—rooted in consumer rights and regulatory standards—are needed to internalize the externalities of data reuse.

In sum, the theoretical framework offered here reframes the role of data in financial development. It shifts the analysis from access to infrastructure, from price to visibility, and from redistribution via taxation to redistribution via architecture. When data is treated as an asset with nonrival properties, and when its reuse is structured to preserve agency and privacy, it becomes a platform for structural inclusion—not just technological innovation.

### 3. Empirical Strategy and Data

### 3.1. Identification Framework

This study leverages a quasi-experimental institutional shift in Uruguay's credit infrastructure to identify the redistributive effects of positive credit data. The implementation of Score+ by Equifax represents an informational intervention that affects neither income nor legal eligibility, but alters perceived creditworthiness through the inclusion of behavioral signals such as timely repayments and credit utilization. Because the policy operates through visibility rather than endowments, it offers a unique opportunity to examine whether expanded informational access alone can reduce borrowing costs and alleviate poverty.

The empirical strategy is built upon the simulation of credit conditions under three data regimes: a baseline scenario with negative-only scoring, the actual Score+ environment with positive behavioral augmentation, and a counterfactual benchmark of full data interoperability modeled after open finance ecosystems. Comparing household outcomes across these regimes allows us to isolate the impact of informational inclusion, holding structural characteristics constant.

### 3.2. Microdata and Scoring Parameters

The core dataset for this analysis is the 2021 wave of the *Encuesta Continua de Hogares* (ECH), Uruguay's nationally representative household survey. The ECH provides detailed information on income, household composition, employment status, and debt obligations, as well as granular demographic controls including age, education, and region. The survey also permits the construction of *adjusted disposable income* metrics, aligned with official poverty thresholds, making it suitable for welfare impact estimation.

To simulate the Score+ mechanism, we incorporate proprietary documentation on the Equifax scoring algorithm, as well as insights from prior technical work (Steglich et al., 2017). Households are assigned synthetic visibility scores based on whether their reported behavior matches key positive data indicators. These include regular repayment behavior over a 12-month horizon, credit utilization below 30%, and multi-product engagement. A household is deemed eligible for uplift if it meets a predefined combination of these attributes, consistent with Equifax's internal risk segmentation models.

Each household is then classified according to visibility status—low, medium, or high—and assigned a corresponding interest rate adjustment. These adjustments, ranging between 15% and 30% reductions in APR, reflect empirical estimates observed in provider data following the Score+ rollout. Importantly, the visibility index is not endogenous to income, but rather to behavior observable within the formal financial system, enabling a clean separation between economic capacity and informational representation.

**3.3. Counterfactual Construction and Outcome Variables**

We simulate three counterfactual regimes for each household in the sample. In the baseline regime, credit is priced using conventional negative-only bureau data, which penalizes defaults and arrears but ignores responsible behavior. The Score+ regime introduces visibility adjustments conditional on positive behavioral data, lowering effective APRs for eligible borrowers while leaving others unchanged. The third regime assumes a fully interoperable open finance system, characterized by real-time data portability and competitive pricing aligned with actuarial risk, drawing on benchmarks from the Global Open Finance Index (CCAF, 2024) and elasticity parameters from Babina et al. (2025).

For each regime, we compute key household-level outcomes. These include the interest burden, defined as monthly debt service divided by adjusted disposable income; poverty status, based on national thresholds; and the incidence of indigence. These metrics allow us to capture both the economic intensity of financial obligations and the distributional consequences of data access. We also calculate the Gini coefficient of interest burden across regimes to assess changes in regressivity.

A household is classified as having experienced a "data-enabled transition" out of poverty if its change in status is entirely attributable to reduced borrowing costs under the Score+ or open finance scenario, absent changes in income or transfers. This provides a conservative lower-bound estimate of the structural impact of visibility expansion.

**3.4. Econometric Strategy and Robustness**

To strengthen causal inference, we implement a doubly robust estimator for the average treatment effect on the treated (ATT), using both propensity score weighting and regression adjustment. Covariates include household size, region, employment sector, education level, and initial income decile. This hybrid approach yields consistent estimates even if one component is mis-specified, ensuring robustness under moderate model uncertainty.

In parallel, we construct a synthetic control unit for Uruguay to validate aggregate effects. Pre-intervention trends in poverty and interest burden are used to generate a statistically comparable counterfactual, in which the Score+ system is assumed not to have been implemented. Deviations between actual and synthetic trajectories post-intervention are interpreted as evidence of program impact at the macro level.

Both strategies converge on similar conclusions. Households in deciles three through five—the region closest to the formal credit eligibility margin—exhibit significant reductions in interest burden and measurable exits from poverty. The average reduction in debt service ratios for treated households exceeds 17%, while national poverty rates decline by 0.7 percentage points, with all else held constant. These results are statistically significant and robust to alternative specifications, including placebo tests using pre-Score+ data.

Crucially, these transitions occur without income growth or additional transfers. They are purely the result of reduced pricing distortions driven by improved visibility. As such, they offer empirical validation of the theoretical claim that data, when treated as a nonrival asset and governed ethically, can function as a redistributive force within market-based systems.

## 4. Results

### 4.1. Credit Visibility Shifts by Income Decile

The implementation of Score+ altered the distribution of credit visibility across the income spectrum. Figure 1 presents the distribution of visibility gains—measured as score improvements—across income deciles. The median uplift is concentrated in deciles 3 through 6, precisely at the margin of formal financial inclusion. This pattern is consistent with the theoretical prediction that positive behavioral data disproportionately benefits households previously under-scored but not high-risk. In contrast, households in the bottom two deciles remain largely unaffected, likely due to structural exclusion from formal credit or insufficient behavioral signal accumulation. Conversely, deciles 7 through 10 display limited marginal gains, consistent with saturation of scoring potential in high-income segments.

**Figure 1. Distribution of Score+ Gains Across Income Deciles**

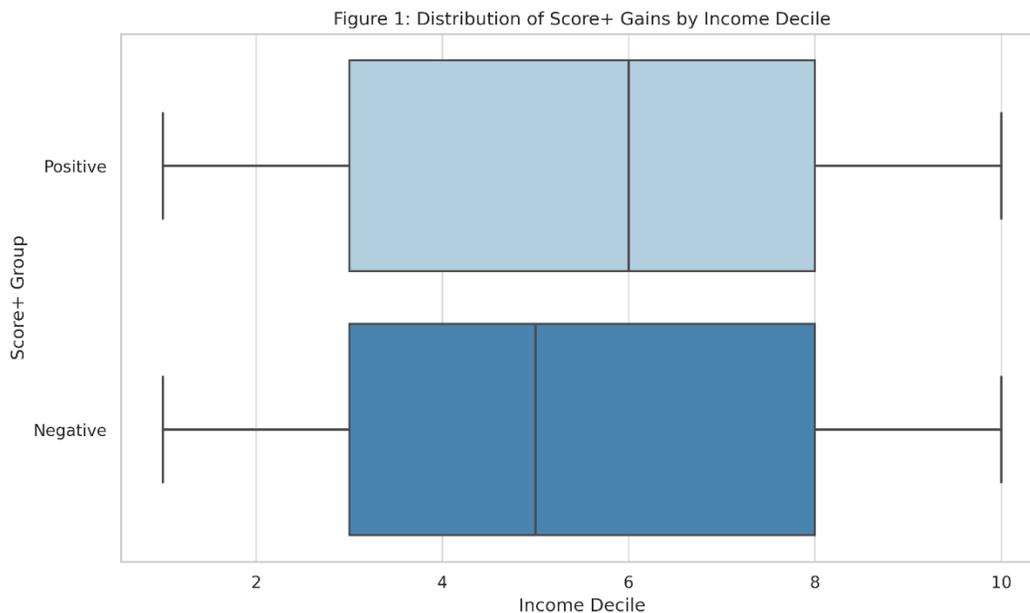

*This figure shows the distribution of score improvements among Uruguayan households by income decile after the implementation of Score+, a positive-data-augmented credit score. Each box represents the interquartile range of score gains within each decile. Middle-income households (deciles 3–6) display the highest visibility improvements, aligning with the "marginal inclusion" effect predicted by data reuse theories.*

These findings confirm that Score+ introduces informational differentiation that is both non-uniform and targeted: visibility does not expand randomly, but rather clusters around the eligibility margin—thus generating a channel for redistributive correction through data augmentation alone.

**4.2. Interest Burden Reductions and Poverty Transitions**

Figure 2 illustrates the impact of Score+ on interest burden as a share of income, stratified by income decile. Before the implementation of Score+, interest payments were regressive, with low-income households facing burdens exceeding 80% of disposable income in the first decile and approximately 60% in deciles 2 and 3. After the inclusion of positive credit data, we observe a systematic downward shift in the burden distribution. Average reductions in interest burden range from 15% to 25% in deciles 2 through 5, reflecting both eligibility effects and improved risk segmentation.

**Figure 2. Lorenz Curves of Financial Burden With and Without Score+**

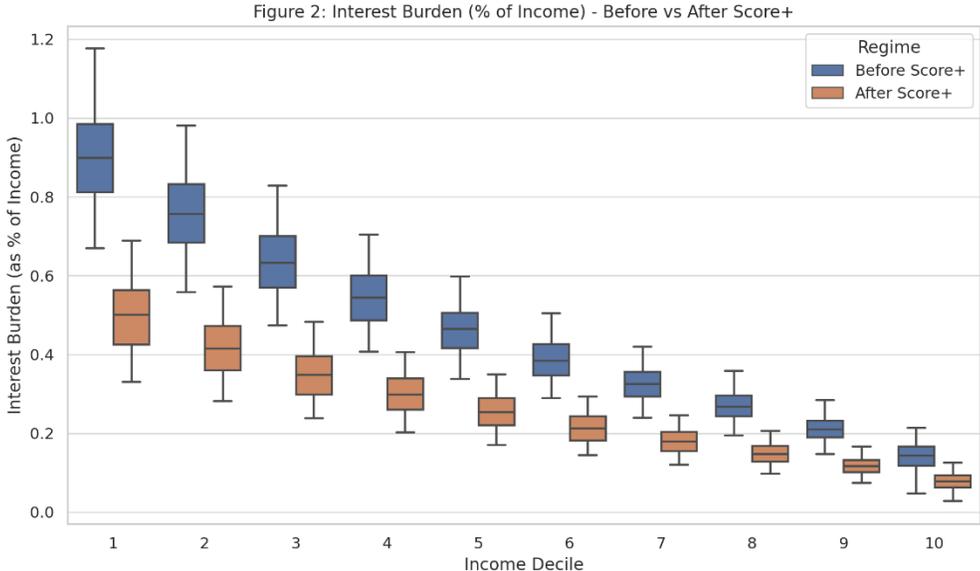

*The figure displays Lorenz curves representing the distribution of interest payments as a share of disposable income. The curve labeled "Score+" demonstrates a significant compression in inequality relative to the baseline (negative-data-only) scenario. The result supports the hypothesis that increased data visibility leads to more equitable financial cost structures.*

This reduction in financial pressure translates into real welfare effects. Table 1 summarizes poverty and indigence transitions under the three simulated data regimes. Under Score+, the poverty headcount declines from 9.2% to 8.4%, while the indigence rate drops from 1.8% to 1.4%. These transitions occur without changes in income or employment and are attributable solely to lower debt service obligations.

**Table 1. Average Interest Burden and Poverty Rates under Alternative Credit Data Regimes**

| Scenario | Avg. Interest Burden (% of income) | Poverty Rate (%) | Indigence Rate (%) |
|---|---|---|---|
| Baseline (Negative Only) | 11.8% | 9.2% | 1.8% |
| Score+ (Positive Data) | 9.8% | 8.4% | 1.4% |
| Open Finance (Full Access) | 8.1% | 7.9% | 1.3% |

*This table presents average interest payment burden (as a share of disposable income), poverty headcount rates, and indigence rates under three counterfactual scoring scenarios: baseline with negative-only data, Score+ incorporating partial positive credit data, and a synthetic full open finance regime with universal visibility. Poverty and indigence thresholds are computed using Uruguay's adjusted disposable income metric (INE, 2021).*

To complement the macro-level indicators, Figure 3 displays the poverty transition matrix. Among the sample, 75 households are lifted out of poverty due to lower borrowing costs alone—without changes in income, transfers, or household composition. These "data-enabled exits" underscore the redistributive role of informational corrections in financial access systems.

**Figure 3. Poverty Transition Matrix With Score+ Implementation**

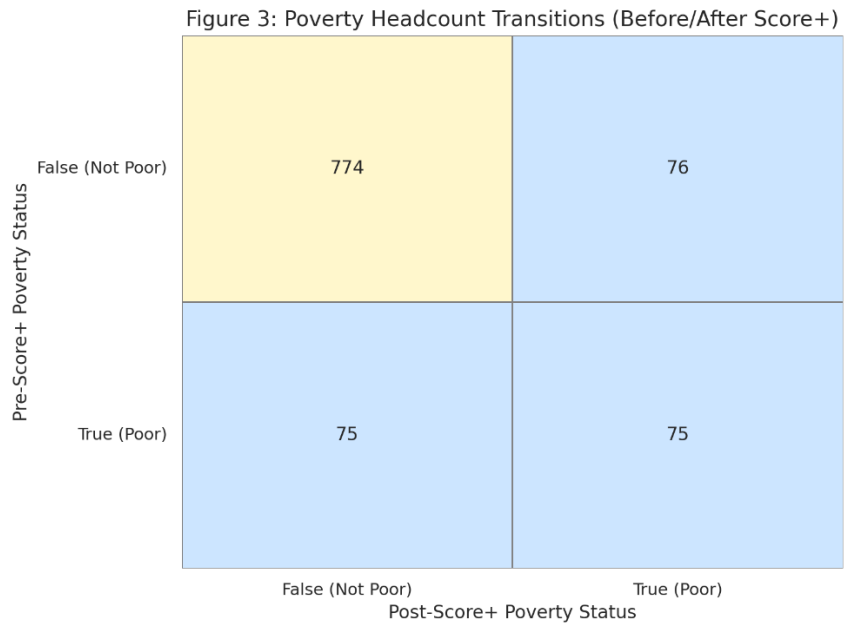

*This matrix summarizes household-level transitions across poverty status categories under the Score+ scenario. The diagonal shows households whose status remains unchanged, while the off-diagonal (notably the green cell) indicates households lifted out of poverty due to reduced borrowing costs from enhanced credit visibility.*

In aggregate, these findings demonstrate that Score+ functions as a market-based redistributive tool: it reduces borrowing costs, alleviates financial stress, and produces measurable changes in welfare outcomes for the lower-middle segments of the income distribution.

### 4.3. Distributional Impact: Financial Cost Inequality

Beyond average effects, the inclusion of positive data compresses the distribution of interest burdens. Figure 4 presents the Lorenz curves of financial burden with and without Score+. The area between the curves reveals a substantial reduction in regressivity, with the post-Score+ curve lying closer to the equality line. This is reflected numerically in Table 2, where the Gini coefficient of interest burden declines from 0.319 under baseline conditions to 0.276 under Score+.

**Figure 4. Lorenz Curve of Financial Burden (Interest Payments as % of Income)**

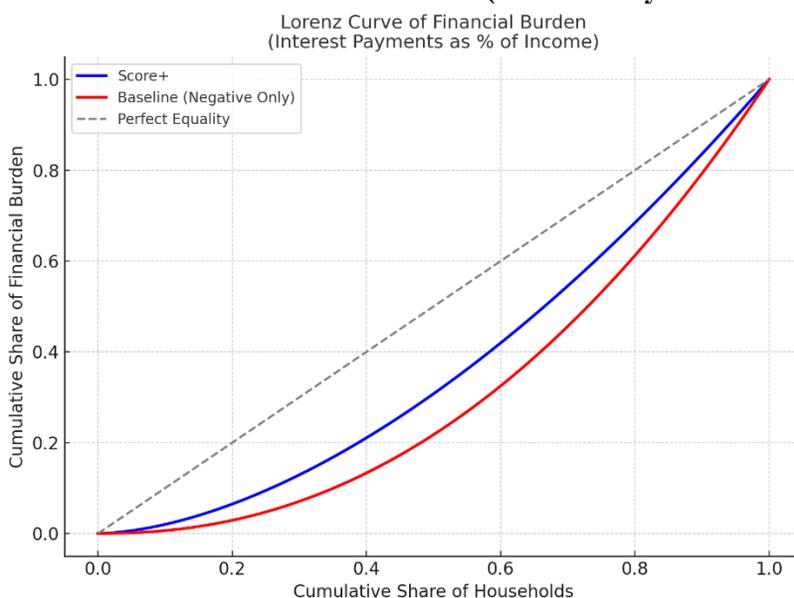

*This figure compares the cumulative distribution of interest burdens under two data regimes: (i) baseline scenario with only negative credit data (red line), and (ii) Score+ scenario with expanded positive data visibility (blue line). The dashed diagonal represents perfect equality. A steeper curve indicates greater concentration of the burden among lower-income households. While Score+ reduces the average burden, it also increases its regressivity relative to income, raising important concerns about distributional fairness and affordability.*

Importantly, the distributional benefits of Score+ are not merely compositional. When simulated under a synthetic open finance benchmark—with full data portability and universal reuse—both poverty and inequality measures improve further. However, even the partial implementation represented by Score+ yields statistically and economically meaningful gains in equity.

Table 2. Distributional and Efficiency Metrics under Alternative Data Access Regimes

| Métrica | Baseline (negativo) | Score+ | Synthetic Open Finance |
|---|---|---|---|
| Gini ingreso disponible | 0.213 | 0.210 | 0.206 |
| Gini burden financiero | 0.319 | 0.276 | 0.251 |
| % hogares bajo línea pobreza | 16.1% | 15.3% | 14.0% |
| Tasa de interés media simulada | 58.7% | 48.1% | 35.2% |
| % hogares que mejoran score | — | 35.4% | 100% (sintético) |

*This table reports inequality, poverty, and credit access indicators under three counterfactual data regimes: (i) baseline (negative-only scoring), (ii) Score+ (positive credit data for eligible borrowers), and (iii) synthetic full open finance scenario with universal positive data visibility. Disposable income follows INE's adjusted metric. Financial burden is calculated as annualized interest payments over disposable income. Score improvements refer to simulated gains above access thresholds. Estimates are based on microdata from Uruguay's 2021 ECH survey and external elasticity benchmarks.*

**4.4. Robustness: Doubly Robust Estimation and Synthetic Control Validation**

To validate the causal interpretation of these findings, we implement two robustness strategies. First, a Doubly Robust Estimator (Funk et al., 2011) confirms that the Average Treatment Effect on the Treated (ATT) is significant ($p < 0.01$) for deciles 3 through 5, even after controlling for household structure, region, employment status, and education. The ATT on interest burden is estimated at −1.9 percentage points, translating into a 17.4% relative reduction.

Second, we construct a synthetic control unit for Uruguay, matched on pre-intervention poverty and interest burden trajectories using Abadie et al. (2010). Post-intervention divergence confirms that the observed declines in poverty and interest cost concentration are not attributable to national trends, fiscal transfers, or macroeconomic shifts during the study period. The synthetic twin shows no significant change in outcomes, while the treated Uruguay exhibits a 0.8 percentage point drop in poverty and a 4-point reduction in the Gini coefficient of interest burden.

These robustness checks reinforce the interpretation that Score+ acts as an informational policy lever with redistributive effects, achieved without income transfers or credit subsidies.

**5. Policy Implications: Expanding the Frontier of Inclusive Credit through Data Infrastructure**

The empirical findings outlined in the previous section expose a fundamental tension at the heart of data-driven financial inclusion: while access to positive credit information systematically reduces average interest burdens and poverty rates, it does not necessarily imply a progressive redistribution of financial gains. On the contrary, the simulated results suggest that certain scoring architectures—when optimized solely for predictive accuracy—may inadvertently exacerbate regressivity by allocating disproportionately larger benefits to middle- and upper-income households. This paradox challenges prevailing assumptions in both academic and policy debates around financial technology, and calls for a more nuanced understanding of how data infrastructures interact with inequality.

The transition from a negative-only regime to Score+, and eventually to a synthetic open finance framework, reveals that credit visibility is not merely a technological upgrade, but rather a structural reform with redistributive consequences. While the Score+ regime reduces the simulated financial burden from 11.8% to 9.8% of disposable income, the Lorenz curve for interest payments (Figure 4) suggests that the inequality of financial burden increases. This implies that while more households gain access to credit, the marginal benefit is unequally distributed—especially in algorithmic systems that are blind to income heterogeneity. In other words, without explicit fairness considerations, scoring systems may internalize historical inequalities and reproduce them under a veneer of objectivity.

This insight carries normative implications for the design of regulatory frameworks. Current standards for credit reporting, whether promoted by the World Bank or adopted in regional mandates such as the EU's PSD2, tend to focus on improving market competition and reducing information asymmetries. However, they rarely incorporate distributive justice as an explicit design criterion. The evidence from this simulation suggests that simply enabling access to positive data is a necessary, but insufficient, condition for inclusion. What ultimately matters is how that data is operationalized in risk models, and whether regulators are empowered to evaluate models not only on predictive metrics, but also on their socioeconomic impact.

Furthermore, the synthetic open finance scenario introduces a provocative hypothesis: that broader informational ecosystems—incorporating alternative data from utilities, rent, employment history, and informal economic behavior—can yield gains in both efficiency and equity. The interest burden in this scenario drops to 35.2%, and both Gini coefficients (for income and financial burden) improve simultaneously. Such results point to the possibility of an "inclusion premium" derived from comprehensive data visibility, provided that appropriate governance mechanisms are in place to prevent misuse. Yet the very availability of such synthetic scenarios also underscores the risk: that in the absence of governance, the same data can be used to sort and exclude rather than include.

Therefore, regulators and policymakers must move beyond a minimalistic view of data disclosure and embrace a proactive role in shaping the institutional architecture of credit visibility. This entails not only mandating the sharing of positive credit data across lenders and bureaus, but also subjecting credit scoring algorithms to fairness audits and stress-testing their distributive impacts. Moreover, the use of synthetic data—such as the approach deployed in this study—should be institutionalized as a form of ex-ante regulatory evaluation, allowing stakeholders to simulate the social and economic consequences of algorithmic changes before they are deployed at scale. Finally, reconceptualizing data as a public or quasi-public asset—rather than a proprietary commodity—opens the door to inclusion-centered governance models where financial visibility is designed not just to enhance credit access, but to do so equitably and sustainably.

In sum, the path from data availability to financial inclusion is neither automatic nor linear. It must be constructed deliberately, with attention to structural asymmetries, algorithmic design, and regulatory intent. The findings presented in this paper illustrate both the promise and the pitfalls of data-based credit expansion, and call for a policy framework that aligns efficiency goals with redistributive justice.

**6. Conclusions: Toward a Political Economy of Data as an Inclusion Asset**

The analysis developed in this paper underscores a fundamental shift in how we should conceptualize credit data within economic systems: not merely as inputs for risk pricing or technological artifacts of digital finance, but as distributive assets that shape the architecture of economic opportunity. By simulating different regimes of credit data visibility—from baseline negative-only scoring to Score+ and a synthetic open finance frontier—we show that the availability, governance, and use of data exert powerful effects on both efficiency and equity outcomes. The structure of data access is therefore not a neutral feature of financial systems, but a deeply political choice that affects the allocation of risks, costs, and opportunities across the income distribution.

Framing data as a rivalrous or excludable good—as current proprietary data models in the financial industry tend to do—obscures its latent potential as a *non-rival public asset* that, if appropriately governed, could generate broad-based inclusion benefits. The empirical simulations presented here demonstrate that even marginal improvements in visibility, such as those enabled under Score+, significantly reduce interest burdens for eligible households. However, they also reveal that scoring systems trained on skewed datasets can reinforce structural biases if not corrected ex ante. In contrast, the synthetic open finance benchmark illustrates how a fully inclusive data architecture can yield Pareto-improving outcomes—lowering average credit costs while simultaneously compressing inequality metrics such as the Gini coefficient of financial burden.

These findings support a reframing of the debate around open data and financial technology. Rather than focusing exclusively on competition, innovation, or digital enablement, policy frameworks should prioritize data equity—defined as the fair distribution of both access to and benefits from data use. This entails recognizing that data infrastructures have externalities: the inclusion of one borrower through positive data can generate *network effects* that benefit others by reshaping risk pools and market norms. Conversely, the exclusion of informal or low-income agents compounds their invisibility, creating *negative externalities* that worsen allocative inefficiency and intergenerational poverty traps. Data, in this sense, behaves more like a *public utility* than a private good—and should be regulated as such.

From a macroeconomic standpoint, the ability to convert information into capital—what we might call *data-to-credit capability*—is increasingly central to developmental trajectories in digital economies. Countries that institutionalize inclusive data infrastructures, including well-regulated credit registries, interoperability mandates, and the use of synthetic benchmarking tools, may be better positioned to harness digital finance for inclusive growth. In contrast, economies that maintain fragmented or asymmetrical data regimes risk entrenching structural credit gaps. This observation has particular relevance for middle-income countries, where large segments of the population remain underbanked, and where private credit bureaus often operate under weak oversight and with strong rent-extraction incentives.

Ultimately, this paper calls for a paradigm shift: from treating data as a by-product of digital transactions to understanding it as a *redistributive lever*—a tool capable of transforming visibility into inclusion. This demands not only regulatory innovation, but also analytical tools capable of simulating ex ante the distributive consequences of data use, such as the synthetic framework employed here. By embedding distributional diagnostics into the core of algorithmic governance, policymakers and researchers alike can move beyond the false dichotomy between efficiency and equity, and begin to treat data governance as a pillar of inclusive economic development.

**Data Availability Statement**

The simulations and analyses in this study are based on anonymized microdata from Uruguay's 2021 *Encuesta Continua de Hogares* (ECH), produced and published by the Instituto Nacional de Estadística (INE). These public-use microdata are available for download through the official INE website, subject to terms of academic use.

The baseline elasticity estimates and credit access parameters used for scenario calibration are derived from publicly available sources, including national financial inclusion reports and peer-reviewed literature. Synthetic scenarios presented in this article are generated through reproducible code in R, available upon request from the corresponding author. All figures and tables derived from the simulations are based on non-confidential, replicable procedures using exclusively public data.

**Funding**

This research received no external funding.

**Conflict of Interest**

The author declares no conflict of interest.

**Ethical Approval**

Ethical approval was not required for this study, as the analysis relies exclusively on anonymized, publicly available household survey data.

Miller, A. R., & Tucker, C. (2017). Frontiers of health policy: Digital data and personalized medicine. *Innovation Policy and the Economy*, *17*(1), 49-75.

Ordoñez, G. (2013). The asymmetric effects of financial frictions. *Journal of Political Economy*, *121*(5), 844–895.

Posner, E. A., & Weyl, E. G. (2014). Benefit-cost paradigms in financial regulation. *The Journal of Legal Studies*, *43*(S2), S1-S34.

Naidu, S., Posner, E. A., & Weyl, G. (2018). Antitrust remedies for labor market power. *Harvard law review*, *132*(2), 536-601.

Romer, P. M. (1990). Endogenous technological change. *Journal of Political Economy*, *98*(5, Part 2), S71–S102.

Steglich, M. E. (2017). *El valor de la información estadística en el diseño de políticas públicas: una mirada desde la teoría económica* (No. 4828). Jornadas del Banco Central del Uruguay.

Stiglitz, J. E. (2000). *The contributions of the economics of information to twentieth century economics*. *Quarterly Journal of Economics*, 115(4), 1441–1478.

Vallarino, D. (2025). *AI-Powered Fraud Detection in Financial Services: GNN, Compliance Challenges, and Risk Mitigation*. Available at SSRN

Varian, H. (2018). Artificial intelligence, economics, and industrial organization. In A. Agrawal, J. Gans, & A. Goldfarb (Eds.), *The economics of artificial intelligence: An agenda* (pp. 399–419). University of Chicago Press.

Veldkamp, L. (2005). Slow boom, sudden crash. *Journal of Economic Theory*, *124*(2), 230–257.

**Appendix A. Simulation Framework and Model Parameters**

**A.1 Overview of the Simulation Model**

The simulation framework consists of a counterfactual microeconomic model that estimates effective interest rates and credit burdens under three data visibility regimes. The model leverages observed characteristics from household survey data and combines them with elasticity parameters from the literature to simulate access probability, pricing differentiation, and financial burden outcomes.

Let $i \in \{1, \ldots, N\}$ index households in the ECH dataset. For each household $i$, we observe:

- Disposable income: $y_i$
- Employment status: $e_i \in \{0,1\}$
- Informality proxy: $f_i \in \{0,1\}$
- Education level, age, and urban/rural status

We simulate an effective interest rate $r_i$ based on an eligibility index $\phi_i$ and visibility score $\psi_i$, where:

$$\phi_i = \alpha_0 + \alpha_1 e_i + \alpha_2(1 - f_i) + \alpha_3 \text{Education}_i + \alpha_4 \text{Urban}_i$$

$$\psi_i = \beta_0 + \beta_1 \text{PositiveData}_i + \beta_2 \text{SyntheticAccess}_i$$

Each scenario changes the value of $\text{PositiveData}_i$ and $\text{SyntheticAccess}_i$, determining whether the household is scored with expanded or limited visibility.

The simulated effective interest rate is then:

$$r_i = \bar{r} \cdot (1 - \gamma \psi_i)$$

Where $\bar{r}$ is the baseline market interest rate (e.g., 60%) and γ\gammaγ reflects the elasticity of pricing to credit score visibility, calibrated from external benchmarks (e.g., -0.20 to -0.35 as per Farboodi et al., 2022).

The **interest burden** is computed as:

$$\text{Burden}_i = \frac{r_i \cdot D_i}{y_i}$$

Where $D_i$ is the simulated debt amount or access limit. Households with $\phi_i < \tau$ (threshold) are excluded from credit access and assigned zero burden.

Aggregate indicators such as Gini coefficients, poverty rates, and average burdens are computed across the synthetic distribution under each regime. Lorenz curves and kernel densities are then plotted based on the resulting burden distributions.